\documentclass[sigconf]{acmart}

\usepackage[utf8]{inputenc}
\usepackage{multirow}
\usepackage{array}
\usepackage{enumitem}

\usepackage{tikz}
\usetikzlibrary{arrows.meta, positioning}

\newcommand{\myparagraph}[1]{\vspace{0.5ex}\noindent \textbf{#1. }}

\renewcommand\footnotetextcopyrightpermission[1]{} %

\setcopyright{none}

\acmConference[SBES 2026]{40th Brazilian Symposium on Software Engineering}{September 8–11, 2026}{São Paulo, SP, Brazil}

\AtBeginDocument{%
  }

\begin{document}

\title[Architectural Pattern Inference via Responsibility Analysis]{From Code to Architecture: Responsibility Analysis and Dependency Graphs in the Inference of Architectural Patterns in Kotlin}

\author{João Pedro Moura}
\email{jpspm@cin.ufpe.br}
\orcid{0000-0002-2536-0956}
\affiliation{%
  \institution{Federal University of Pernambuco}
  \city{Recife}
  \state{Pernambuco}
  \country{Brazil}
}

\author{Breno Miranda}
\email{bafm@cin.ufpe.br}
\orcid{0000-0001-9608-9393}
\affiliation{%
  \institution{Federal University of Pernambuco}
  \city{Recife}
  \state{Pernambuco}
  \country{Brazil}
}

\renewcommand{\shortauthors}{Moura and Miranda}

\begin{abstract}
Understanding software architecture is essential for maintenance and evolution. However, architectural decisions are often undocumented or outdated, forcing developers to infer architectural structures directly from source code. This task is challenging due to inconsistent naming, variable code quality, architectural erosion, hybrid styles, and the lack of systematic identification methods.
In this work, we propose an automated pipeline to infer architectural patterns directly from Kotlin source code. The approach combines static code analysis, heuristic identification of class responsibilities, and dependency graph construction to capture both the roles of individual components and the structural relationships between them. Based on this representation, the pipeline is able to identify architectural patterns such as MVVM, MVP, MVC, MVI, and VIPER. To evaluate the effectiveness of the proposed approach, we conducted an empirical study using a dataset of Kotlin repositories collected from GitHub. The evaluation was performed on a statistically representative sample, comparing the inferred architectures with those explicitly declared by the repository authors. The results show that the pipeline achieves an accuracy of 69.8\%, significantly outperforming benchmark strategies such as random and majority class classifiers.
These results indicate that combining responsibility analysis with dependency-based structural representations provides a robust and scalable solution for inferring architectural patterns. The proposed approach can aid in software maintenance, architectural analysis, and large-scale empirical studies in modern software ecosystems.
\end{abstract}

\keywords{Mobile Applications, Software Maintenance, Reengineering, and Evolution, Software Archtecture}

\maketitle

\section{Introduction}

Software systems are increasingly complex and continuously evolving, making it difficult to maintain a clear understanding of their underlying architecture. In many cases, architectural decisions are poorly documented or become outdated, requiring developers to infer the system’s structure directly from source code. This problem is further aggravated by architectural erosion, which degrades structural integrity over time and negatively impacts software quality and evolution~\cite{https://doi.org/10.1002/smr.2423}.

Architectural patterns are widely adopted to structure software systems, particularly in mobile development. However, different patterns often share similar components and interaction mechanisms, making their identification non-trivial when analyzing real-world codebases. This challenge is well recognized in practice, where even novice developers struggle to correctly identify architectural patterns and lack systematic methods or tool support for this task~\cite{stol2010identifying}.

To address this problem, this work proposes an automated pipeline for inferring architectural patterns directly from Kotlin source code. The approach combines static code analysis, heuristic identification of class responsibilities, and dependency graph construction to capture both component roles and their relationships.

The main contributions of this work are: (i) a pipeline that integrates responsibility analysis and dependency graphs for architectural inference; (ii) an approach that identifies architectural patterns from implementation artifacts without relying solely on naming conventions or supervised learning; and (iii) an empirical evaluation on Kotlin repositories demonstrating the effectiveness of the proposed method.

The remainder of this paper is organized as follows. Section~\ref{sec:context} presents the context and motivation. Section~\ref{sec:related} discusses related work. Section~\ref{sec:methodology} describes the proposed methodology. Section\ref{sec:evaluation} presents the evaluation setup, followed by the analysis of the research questions in Section~\ref{sec:rqs}. Section~\ref{sec:discussion} Discusses the overall performance, Section~\ref{sec:threats} discusses threats to validity, and Section~\ref{sec:conclusion} concludes the paper.

\section{Context and Motivation}
\label{sec:context}

As software systems evolve, maintaining a clear understanding of their architecture becomes increasingly difficult. Architectural decisions are often undocumented or become outdated, forcing developers to infer architecture directly from source code.

This activity becomes even more challenging due to poor development practices, inconsistent naming conventions, the existence of hybrid architectures, and the reuse of similar elements in various architectural styles.

Therefore, automatically detecting architectural patterns from source code represents a complex challenge, as it requires not only identifying the functions of individual classes but also understanding the relationships and interaction processes between them. Furthermore, ensuring the existence of specific components, such as ViewModels or Presenters, does not necessarily indicate a specific architecture, since similar components can be found in various architectural patterns.

Therefore, systematic approaches are needed to analyze software systems and infer their architectural patterns. In this context, this work proposes a workflow to identify class functions, analyze dependencies, and deduce these patterns based on their relationships.

\myparagraph{Practical Importance}
Automated identification of architectural patterns has significant implications for both software practice and research. In practice, it aids in maintenance, developer integration, and compliance verification. Analytically, it facilitates repository mining, analysis of software evolution, and large-scale empirical studies.

\myparagraph{Design Patterns}
Design patterns describe reusable object-oriented solutions to recurring design problems. Unlike architectural patterns, they focus on object interactions rather than the overall organization of a software system ~\cite{explainer_pattern} ~\cite{design_patterns_abstraction}. In this work, design patterns are discussed only to distinguish them from architectural patterns.

\myparagraph{Architecture Patterns}
Architectural patterns have become conventional solutions in software development, impacting not only the efficiency of the development process but also the long-term performance of the software, including its maintenance and the addition of new features \cite{ilhamevaluating}.

Software architecture lays the foundation for subsequent phases of the software development lifecycle, assisting developers in the correct development of software. An architectural pattern is a well-established structural scheme for software systems that affects the basic structure of the system. It allows for the documentation of architectural design decisions, the facilitation of connection between stakeholders through a typical dictionary, and the analysis of software quality aspects.

Although several architectural patterns are available, the developer must choose the one that is most appropriate for the application in question.

Software architects use both architectural patterns and strategies in their work. Patterns provide the structure and behavior of software systems as a means of satisfying numerous system requirements. Strategies are design decisions that address specific quality attribute issues \cite{software_design_pattern}.

In this work, we focus on a set of architectural patterns widely adopted in software development, called MVC, MVP, MVVM, MVI, and VIPER.

These patterns were selected due to their widespread adoption, especially in mobile and frontend development, as well as their clear separation of responsibilities, which makes them suitable for role-based analysis.

\begin{itemize}[leftmargin=*]
        \item \textbf{MVVM (Model-View-ViewModel):}
       uses a ViewModel to expose state and behavior to the View, often leveraging data-binding mechanisms to synchronize UI and data.
        \item \textbf{MVP (Model-View-Presenter):}
        introduces a Presenter component that replaces the Controller, promoting a more explicit separation between the View and the business logic.
        \item \textbf{MVC (Model-View-Controller):}
        separates the system into three main components: the Model, responsible for data and business logic; the View, responsible for presentation; and the Controller, which mediates interactions between them.
        \item \textbf{MVI (Model-View-Intent):}
       is based on a unidirectional data flow, where user intents are transformed into states that are rendered by the View.
        \item \textbf{VIPER (View-Interactor-Presenter-Entity-Router):}
        decomposes the system into highly modular components, improving testability and separation of concerns at the cost of increased complexity.
    \end{itemize}

Although these patterns define distinct architectural structures, they often share similar components and naming conventions. For example, elements such as ViewModels, Presenters, or Controllers may appear in multiple patterns, leading to ambiguity when trying to identify the underlying architecture based solely on class names.

Therefore, distinguishing between architectural patterns requires analyzing not only the presence of specific functions but also the relationships and interaction flows between components.

\myparagraph{Why Kotlin?}
Kotlin has emerged as one of the dominant programming languages for Android and cross-platform development. Its expressive syntax, strong interoperability with Java, and official Google support have driven its widespread adoption in both industry and open-source communities. Furthermore, Kotlin applications frequently employ modern architectural patterns such as MVVM, MVI, and VIPER, making the language an ideal target for empirical architectural studies. Despite this growing relevance, Kotlin-specific tools for automated architectural inference remain scarce.

Kotlin has also received significant investment in its cross-platform framework, with support from JetBrains and Google. Kotlin Cross-Platform (KMP) has been adopted by companies such as Google, Duolingo, Forbes, Philips, McDonald's, Bolt, H\&M, Baidu, Kuaishou, and Bilibili due to the increased performance and quality reported by 65\% of teams, according to a KMP survey from the second half of 2024. In addition to native performance, it generates native binaries and accesses platform APIs directly in situations where virtual machines are undesirable or impossible, such as on iOS \cite{jetbrains_kmp_overview}.

\section{Related Work}
\label{sec:related}

Empirical studies on software architecture increasingly depend on the availability of structured datasets and reproducible benchmarks. Such resources enable researchers to evaluate proposed techniques, compare alternative approaches, and conduct large-scale analyses across software ecosystems. Over the past years, several initiatives have addressed this need, targeting different aspects of architectural analysis, sustainability, and pattern detection. However, most existing datasets remain limited in scope, often focusing on textual artifacts, specific domains, or a small subset of architectural styles.

Afina Et al. \cite{11430736} conduct an empirical study that evaluates the use of Kotlin Multiplatform in conjunction with the Model-View-Intent architecture in cross-platform mobile development, comparing this approach with native Android implementations and with Flutter, based on equivalent applications on the different platforms and analyzing runtime performance metrics such as startup time and memory consumption.

To advance research on sustainable software design, Ahmadisakha et al.~\cite{10.1007/978-3-031-71246-3_14} investigated practitioner discussions on online platforms and combined these findings with repository mining to construct annotated datasets centered on sustainability-related architectural concerns. Their study revealed that practitioners primarily emphasize technical and economic dimensions, particularly design and implementation, while paying comparatively less attention to evaluation and maintenance activities. Although this work provides valuable insights into architectural reasoning and decision-making, its focus lies on textual discussions and developer perceptions rather than on the structural characteristics of implemented software systems.

Other efforts have sought to improve reproducibility in software architecture research by developing benchmark datasets for specific tasks. For instance, Corallo et al.~\cite{10.1007/978-3-031-36889-9_30} proposed a benchmark to evaluate information retrieval techniques for recovering traceability links between architectural documentation and software models. This contribution established an important baseline for reproducible experimentation in architecture recovery. Nevertheless, the benchmark concentrates on documentation artifacts and model traceability, without explicitly representing architectural structures extracted from source code.

Closer to the problem addressed in this work, Komolov et al.~\cite{komolov2022towards} applied machine learning to classify architectural patterns to identify architectural patterns such as MVP and MVVM from source code metrics and manually labeled repositories. Their results showed that automated architectural classification is feasible. However, their approach was limited to a small number of architectural styles and depended heavily on supervised learning, which may restrict generalizability across languages, frameworks, and emerging architectural paradigms.

Extending this line of work, Singh et al. \cite{11308595} It proposes an Artificial Intelligence-based design pattern recommendation system to support architectural decisions and automate refactorings during software development. The approach combines multimodal source code representations, Graph Neural Networks (GNNs), Transformer models, knowledge graphs, and self-supervised learning to identify design patterns, detect anti-patterns, and recommend refactorings suited to the software context. Furthermore, the system incorporates mechanisms for the automatic synthesis and verification of refactorings, utilizing compilation, regression testing, and static analysis to validate the proposed modifications.

Other efforts to pattern detection have been made. Fontana et al.~\cite{ARCELLIFONTANA20111306} propose the MARPLE tool, a plug-in developed for the Eclipse platform, with the aim of supporting reverse engineering activities, especially the detection of design patterns and the reconstruction of software architecture. The approach is based on the automatic extraction of basic elements and metrics directly from the source code, using libraries for data access and graph manipulation. The article focuses mainly on the process of identifying design patterns, seeking to help developers better understand the structure and organization of existing systems through abstractions at different levels.

Collectively, these studies have made important contributions to the empirical study of software architecture. They have improved reproducibility, demonstrated the feasibility of automated pattern detection, and highlighted the value of structured architectural data. At the same time, they also expose a significant gap: the absence of large-scale, publicly available datasets that capture architectural patterns directly from implementation artifacts across modern software ecosystems.

This gap is particularly pronounced in the Kotlin ecosystem. Despite the widespread adoption of Kotlin in Android and cross-platform development, few empirical resources focus specifically on its architectural landscape. Modern Kotlin applications frequently employ sophisticated patterns such as MVVM, MVI, and VIPER, often combined with frameworks like Jetpack Compose, Coroutines, Flow, and Hilt. These technologies introduce structural nuances that are not adequately represented in existing datasets derived from other languages or older architectural paradigms.

The lack of architectural datasets limits researchers' ability to conduct large-scale empirical studies, evaluate architectural inference techniques, and investigate the evolution of architectural practices in modern software. Overcoming this limitation is essential for advancing architecture mining, automated pattern detection, and evidence-based software engineering research in one of today's most relevant development ecosystems.

\section{Methodology}
\label{sec:methodology}

\subsection{Proposed Solution}

To address these challenges, this work proposes an automated pipeline to infer architectural patterns directly from source code. The approach combines static code analysis, heuristic identification of class roles, and dependency graph analysis to capture both the responsibilities of individual components and the relationships between them.

By evaluating the resulting dependency structures against characteristic architectural motifs, the proposed pipeline can automatically identify patterns such as MVVM, MVP, MVC, MVI, and VIPER. This enables scalable and reliable architectural classification across large collections of Kotlin repositories, supporting both practical software engineering tasks and empirical research.

Although this work focuses on Kotlin repositories, the overall workflow is language-independent in principle. Adapting the pipeline to another language requires redefining the language-specific heuristics used for role identification and architectural motif extraction.

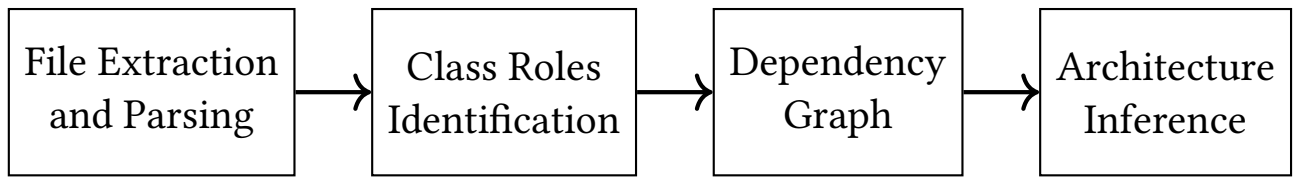
\begin{figure}[ht]
\centering
\begin{tikzpicture}[
    node distance=0.5cm,
    every node/.style={
        draw,
        rectangle,
        minimum width=1.1cm,
        minimum height=1.1cm,
        align=center,
        font=\small
    },
    arrow/.style={->, thick}
]

\node (A) {File Extraction\\and Parsing};
\node (B) [right=of A] {Class Roles\\Identification};
\node (C) [right=of B] {Dependency\\Graph};
\node (D) [right=of C] {Architecture\\Inference};

\draw[arrow] (A) -- (B);
\draw[arrow] (B) -- (C);
\draw[arrow] (C) -- (D);

\end{tikzpicture}
\caption{Pipeline Overview}
\label{fig:overview}
\end{figure}

\subsection{File Extraction and Parsing}

The first step of the proposed pipeline shown in Figure \ref{fig:overview} consists of scanning recursively to extract and parse Kotlin source files from the repository.

To analyze only production code, several filtering heuristics are applied. Files located in directories typically associated with non-production artifacts, such as test folders, examples, templates, benchmarks, generated files, and build infrastructure, are ignored to prevent noise in the architectural inference.

Using the Tree-sitter with the Kotlin grammar, it generates an abstract syntax tree (AST) for each file, which allows the extraction of structural information to use them to determine the architectural role of each class and to identify dependencies among them.
The information extracted includes \textit{package declarations}, \textit{import statements}, \textit{inheritance relationships}, \textit{annotations}, and \textit{referenced symbols}.

\subsection{Class Roles Identification}

After extracting structural information from the source code, the next step consists of identifying the architectural role of each class. This classification is performed using role-specific heuristic rules that capture common naming conventions, structural patterns, framework annotations, inheritance relationships, and package organization used in Kotlin applications.

The role identification process combines multiple sources of evidence:

Each source proposes candidate roles for the analyzed class. The candidate roles are then aggregated before applying a deterministic precedence strategy to resolve conflicts and determine a single architectural role.
Explicit semantic indicators, such as annotations and inheritance relationships, are prioritized over structural indicators derived from package organization, imports, and naming conventions. When multiple roles remain possible, a predefined role resolution order is applied to ensure a unique architectural role assignment.

\myparagraph{Naming Conventions}
Using tokenization techniques to split camelCase, the class names are analyzed to detect common architectural terms such as:
\textit{View}, \textit{ViewModel}, \textit{Presenter}, \textit{Controller}, \textit{Interactor}, \textit{UseCase}, \textit{Repository}, \textit{Service}, \textit{Entity}, \textit{Model}, \textit{Intent}, \textit{State}, \textit{Router}.

\myparagraph{Annotations}
Annotations commonly associated with architectural layers are also considered, for example:
\begin{itemize}
    \item \textit{@Composable} indicates a UI component
    \item \textit{@HiltViewModel} suggests a ViewModel
    \item \textit{@Serializable} suggests a data model
    \item \textit{@Module} and \textit{@Provides} indicate DI components
\end{itemize}

For example, a class annotated with @HiltViewModel, extending ViewModel, or located under presentation/viewmodel accumulates evidence for the ViewModel role from different heuristic sources.

\myparagraph{Imports}
Imported framework classes provide additional signals. For instance, classes importing \textit{androidx.lifecycle.ViewModel} accumulate evidence for the ViewModel role, while imports from \textit{retrofit2} or \textit{ktor} may indicate service or network layers.

\myparagraph{Inheritance}
Inheritance relationships are also analyzed. Classes extending known base types such as \textit{ViewModel}, \textit{ComponentActivity} or \textit{Fragment} provide strong evidence of their architectural role.

\myparagraph{Package and URI structure}
Folder structure conventions are leveraged to detect roles based on common project layouts, such as presentation/viewmodel for ViewModels or domain/usecase for UseCases.

After aggregating the candidate roles produced by all sources, the final architectural role is determined according to a predefined role resolution order that prioritizes architecture-defining roles whenever multiple candidate roles remain.

Classes that represent auxiliary utilities, dependency injection configuration, or test infrastructure are excluded to prevent noise from entering the architectural analysis.

\subsection{Dependency Graph}

    After class roles are identified, the system constructs a dependency graph, which represents the structural relationships between them.

    Dependencies are inferred from three main sources extracted during parsing:

    \begin{enumerate}
        \item \textbf{Inheritance relationships:}
        If a class extends or implements another class or interface, a dependency edge is created
        \item \textbf{Import statements:}
        If a class imports another class defined in the project, this indicates a potential dependency.
        \item \textbf{Symbol references:}
        References to other class names within the source code are also considered as dependencies.
    \end{enumerate}
    To focus on architecture-relevant relationships, the graph includes only classes whose roles correspond to architectural components such as UI, presentation logic, domain interaction, or domain data.

    Furthermore, architectural constraints are applied to ensure that only meaningful relationships are preserved. For example, certain roles are allowed to depend on others according to common architectural layering principles (e.g., views may depend on ViewModels, but entities should not depend on UI components).

    The graph is pruned by removing weak and redundant connections. A pruning step limits the number of outgoing dependencies per node and removes isolated nodes that do not participate in the architectural structure. This produces a focused dependency graph representing the core architectural relationships of the application.

    The dependency graph is not directly classified. Instead, it serves as the structural representation from which architectural dependency motifs are extracted. These motifs capture characteristic interactions between architectural roles and constitute the primary evidence used during architecture inference.

    In this work, a motif corresponds to a directed dependency between architectural roles (e.g., \textit{View} $\rightarrow$ \textit{ViewModel}). More complex architectural structures emerge from the combination of multiple motifs within the dependency graph.

\subsection{Architecture Inference}

    The final step of the pipeline consists of inferring the architectural patterns implemented in the analyzed project. The approach evaluates candidate architectures, including

    \begin{itemize}[leftmargin=*]
        \item \textbf{MVVM:}
        dependencies from View -> ViewModel and ViewModel -> Domain components.
        \item \textbf{MVP:}
        bidirectional interaction between View and Presenter, with Presenter accessing domain data.
        \item \textbf{MVC:}
        View interacting with a controller that coordinates access to domain data.
        \item \textbf{MVI:}
        unidirectional event flow involving intents, states, and domain interactions.
        \item \textbf{VIPER:}
        collaboration between presenter, interactor, router, and entity layers.
    \end{itemize}

    Architecture inference is performed through a multi-stage decision process rather than a simple comparison of scores. First, architectural motifs extracted from the dependency graph are compared against the expected motifs for each supported architectural pattern. Each detected motif contributes to the score of the architectural patterns in which it is expected. The architectures with the highest scores are then identified as candidate solutions.

When a single architecture achieves the highest score, it is selected directly. When multiple architectures obtain the same highest score, ties are resolved using the architectural roles previously assigned during class role identification. For example, the presence of architecture-specific roles (such as Interactor or Router) is used to distinguish between patterns that exhibit similar dependency structures. If the tie-breaking strategy fails to unequivocally identify a dominant architecture, the repository is classified as "Unknown."

Finally, a confidence value is calculated by normalizing the score of the selected architecture with the total evidence accumulated across all candidate architectures. This confidence index provides an estimate of how strongly the observed architectural motifs support the inferred classification.

\section{Evaluation}
\label{sec:evaluation}

    \subsection{Experimental Subjects}

    We evaluated the pipeline using a large-scale collection of Kotlin repositories was obtained from GitHub using its search API. Repositories were retrieved using architecture-specific queries targeting the five architectural patterns considered in this study:
    \textit{language:kotlin topic:mvvm; language:kotlin topic:mvp; language:kotlin topic:mvi; language:kotlin topic:mvc; language:kotlin topic:viper;}

These queries were designed to maximize the probability of retrieving repositories explicitly associated with each architectural style. A total of 2,000 repositories were initially collected.

Subsequently, a filtering step was applied to ensure legal availability and suitability for analysis. Repositories without an explicit open-source license were excluded, resulting in a reduced set of 733 repositories.

Since the goal of the evaluation is to compare the architecture inferred by the pipeline with a reliable reference label, an additional manual screening was performed. Each repository was inspected to determine if its architectural pattern was explicitly declared by its authors in at least one of the following sources:
i) \textit{Repository Title}, ii) \textit{Repository Description}, or iii) \textit{README Documentation}.

Only repositories that provided an explicit architectural declaration were retained for validation. This criterion minimizes ambiguity in reference labels and establishes a reliable fundamental truth for the empirical evaluation.

After this final selection step, the validation population consisted of 527 Kotlin repositories encompassing the five architectural patterns considered in this work. This selected set served as the basis for sample selection and subsequent experimental evaluation.

To evaluate the effectiveness of the proposed pipeline, this study aims to answer the following research questions:

\begin{itemize}

\item \textbf{RQ1:} How accurately does the proposed pipeline infer architectural patterns in Kotlin repositories?

\item \textbf{RQ2:} How does the proposed pipeline compare to existing architectural classification approaches?

\item \textbf{RQ3:} How does the proposed pipeline compare to a Large Language Model (LLM) in inferring architectural patterns?

\end{itemize}

\begin{table*}[htbp]
\centering
\caption{Validation Dataset Summary}
\label{tab:validation-dataset}
\scriptsize
\begin{minipage}{0.48\textwidth}
\centering
\begin{tabular}{r p{5.8cm} rr}
\toprule
\# & Repository & Classes & Deps \\
\midrule
1 & \href{https://github.com/skydoves/Pokedex}{Pokedex} & 34 & 8 \\
2 & \href{https://github.com/ahmedeltaher/Android-MVVM-Architecture}{Android-MVVM-Architecture} & 45 & 17 \\
3 & \href{https://github.com/PatilShreyas/Foodium}{Foodium} & 25 & 6 \\
4 & \href{https://github.com/VIPyinzhiwei/Eyepetizer}{Eyepetizer} & 183 & 86 \\
5 & \href{https://github.com/mhss1/MyBrain}{MyBrain} & 351 & 251 \\
6 & \href{https://github.com/zhujiang521/PlayAndroid}{PlayAndroid} & 132 & 66 \\
7 & \href{https://github.com/skydoves/DisneyMotions}{DisneyMotions} & 25 & 3 \\
8 & \href{https://github.com/lulululbj/wanandroid}{wanandroid} & 56 & 31 \\
9 & \href{https://github.com/skydoves/MarvelHeroes}{MarvelHeroes} & 22 & 3 \\
10 & \href{https://github.com/skydoves/pokedex-compose}{pokedex-compose} & 59 & 30 \\
11 & \href{https://github.com/skydoves/DisneyCompose}{DisneyCompose} & 15 & 3 \\
12 & \href{https://github.com/Drjacky/MVVMTemplate}{MVVMTemplate} & 55 & 19 \\
13 & \href{https://github.com/theapache64/stackzy}{stackzy} & 93 & 56 \\
14 & \href{https://github.com/dbaroncelli/D-KMP-sample}{D-KMP-sample} & 31 & 28 \\
15 & \href{https://github.com/EranBoudjnah/CleanArchitectureForAndroid}{CleanArchitectureForAndroid} & 157 & 37 \\
16 & \href{https://github.com/odaridavid/Clean-MVVM-ArchComponents}{Clean-MVVM-ArchComponents} & 65 & 29 \\
17 & \href{https://github.com/ffgiraldez/reactive-mvvm-android}{reactive-mvvm-android} & 47 & 19 \\
18 & \href{https://github.com/mars885/gamedge}{gamedge} & 456 & 329 \\
19 & \href{https://github.com/skydoves/Pokedex-AR}{Pokedex-AR} & 29 & 10 \\
20 & \href{https://github.com/furkanaskin/Weatherapp}{Weatherapp} & 83 & 56 \\
21 & \href{https://github.com/Joker-x-dev/CoolMallKotlin}{CoolMallKotlin} & 250 & 166 \\
22 & \href{https://github.com/mobiledevpro/Android-Kotlin-MVVM-Template}{Android-Kotlin-MVVM-Template} & 46 & 19 \\
23 & \href{https://github.com/fuusy/component-jetpack-mvvm}{component-jetpack-mvvm} & 97 & 23 \\
24 & \href{https://github.com/skydoves/TheMovies}{TheMovies} & 89 & 26 \\
25 & \href{https://github.com/zhujiang521/PlayWeather}{PlayWeather} & 59 & 15 \\
26 & \href{https://github.com/skydoves/TheMovies2}{TheMovies2} & 60 & 16 \\
27 & \href{https://github.com/imashnake0/Animite}{Animite} & 76 & 10 \\
28 & \href{https://github.com/gastsail/CocktailApp}{CocktailApp} & 29 & 15 \\
29 & \href{https://github.com/TakuSemba/JetHub}{JetHub} & 64 & 15 \\
30 & \href{https://github.com/jenly1314/MVVMFrame}{MVVMFrame} & 62 & 12 \\
31 & \href{https://github.com/zedlabs/WallPortal}{WallPortal} & 28 & 7 \\
32 & \href{https://github.com/wajahatkarim3/Imagine}{Imagine} & 43 & 9 \\
33 & \href{https://github.com/GetStream/AvengersChat}{AvengersChat} & 59 & 18 \\
34 & \href{https://github.com/mertceyhan/bitcoin-market-android}{bitcoin-market-android} & 36 & 20 \\
35 & \href{https://github.com/skydoves/GoldMovies}{GoldMovies} & 91 & 20 \\
36 & \href{https://github.com/Kotlin-Android-Open-Source/Jetpack-Compose-MVI-Coroutines-Flow}{Jetpack-Compose-MVI-Coroutines-Flow} & 77 & 63 \\
37 & \href{https://github.com/DanielRendox/RoutineTracker}{RoutineTracker} & 125 & 64 \\
38 & \href{https://github.com/happysingh23828/Android-Clean-Architecture}{Android-Clean-Architecture} & 76 & 45 \\
39 & \href{https://github.com/skydoves/GithubFollows}{GithubFollows} & 51 & 15 \\
40 & \href{https://github.com/hongbeomi/HarryPotter}{HarryPotter} & 17 & 2 \\
41 & \href{https://github.com/cdalwyn/PlayAndroid}{PlayAndroid} & 211 & 146 \\
42 & \href{https://github.com/idisfkj/AwesomeGithub}{AwesomeGithub} & 136 & 69 \\
43 & \href{https://github.com/CoinTrend/CoinTrend}{CoinTrend} & 168 & 98 \\
44 & \href{https://github.com/ErickSumargo/Dads}{Dads} & 121 & 22 \\
45 & \href{https://github.com/kaushikgopal/movies-usf-android}{movies-usf-android} & 37 & 19 \\
46 & \href{https://github.com/UCodeUStory/Relax}{Relax} & 145 & 26 \\
47 & \href{https://github.com/theapache64/topcorn}{topcorn} & 36 & 7 \\
48 & \href{https://github.com/auron567/Gallerit}{Gallerit} & 33 & 8 \\
\bottomrule
\end{tabular}
\end{minipage}
\hfill
\begin{minipage}{0.48\textwidth}
\centering
\begin{tabular}{r p{5.8cm} rr}
\toprule
\# & Repository & Classes & Deps \\
\midrule
49 & \href{https://github.com/akoufa/CoolWeather}{CoolWeather} & 56 & 30 \\
50 & \href{https://github.com/haroldadmin/Vector}{Vector} & 56 & 24 \\
51 & \href{https://github.com/dgngulcan/droid-feed}{droid-feed} & 92 & 65 \\
52 & \href{https://github.com/YummyLau/AndroidModularArchiteture}{AndroidModularArchiteture} & 254 & 23 \\
53 & \href{https://github.com/wasabeef/kotlin-mvvm}{kotlin-mvvm} & 43 & 5 \\
54 & \href{https://github.com/anitab-org/mentorship-android}{mentorship-android} & 82 & 29 \\
55 & \href{https://github.com/charlee-dev/PexWallpapers}{PexWallpapers} & 61 & 17 \\
56 & \href{https://github.com/igorwojda/android-showcase}{android-showcase} & 74 & 35 \\
57 & \href{https://github.com/SkyD666/PodAura}{PodAura} & 771 & 413 \\
58 & \href{https://github.com/Kotlin-Android-Open-Source/MVI-Coroutines-Flow}{MVI-Coroutines-Flow} & 85 & 61 \\
59 & \href{https://github.com/respawn-app/FlowMVI}{FlowMVI} & 270 & 186 \\
60 & \href{https://github.com/badoo/MVIKotlin}{MVIKotlin} & 191 & 32 \\
61 & \href{https://github.com/yogeshpaliyal/KeyPass}{KeyPass} & 93 & 46 \\
62 & \href{https://github.com/vishal2376/snaptick}{snaptick} & 106 & 97 \\
63 & \href{https://github.com/v1tzor/TimePlanner}{TimePlanner} & 501 & 598 \\
64 & \href{https://github.com/Ezike/Baking-App-Kotlin}{Baking-App-Kotlin} & 149 & 164 \\
65 & \href{https://github.com/MoIbrahim15/Android-Kotlin-MVI-CleanArchitecture}{Android-Kotlin-MVI-CleanArchitecture} & 143 & 109 \\
66 & \href{https://github.com/hoc081098/ComicReaderApp_MVI_Coroutine_RxKotlin_Jetpack}{ComicReaderApp} & 264 & 147 \\
67 & \href{https://github.com/Ezike/StarWarsSearch-MVI}{StarWarsSearch-MVI} & 135 & 112 \\
68 & \href{https://github.com/odaridavid/WeatherApp}{WeatherApp} & 59 & 37 \\
69 & \href{https://github.com/arthur3486/android-mvvm}{android-mvvm} & 52 & 16 \\
70 & \href{https://github.com/gabriel-TheCode/Dagger-Hilt-MVVM}{Dagger-Hilt-MVVM} & 20 & 10 \\
71 & \href{https://github.com/kaleidot725/Jetpack-Compose-Orbit-MVI-Demo}{Orbit-MVI-Demo} & 42 & 17 \\
72 & \href{https://github.com/memtrip/android-mvi}{android-mvi} & 173 & 186 \\
73 & \href{https://github.com/frankois944/kmp-mvvm-exploration}{kmp-mvvm-exploration} & 42 & 22 \\
74 & \href{https://github.com/git-xuhao/KotlinMvp}{KotlinMvp} & 116 & 26 \\
75 & \href{https://github.com/hoc081098/WeatherApp_MVI_sample}{WeatherApp\_MVI\_sample} & 123 & 35 \\
76 & \href{https://github.com/iceCola7/KotlinMVPSamples}{KotlinMVPSamples} & 37 & 8 \\
77 & \href{https://github.com/marcherdiego/android_mvp}{android\_mvp} & 53 & 15 \\
78 & \href{https://github.com/pangli/KotlinMvp}{KotlinMvp} & 40 & 7 \\
79 & \href{https://github.com/easyandroidgroup/EasyAndroid}{EasyAndroid} & 46 & 1 \\
80 & \href{https://github.com/TonnyL/PaperPlane}{PaperPlane} & 113 & 8 \\
81 & \href{https://github.com/iOrchid/android-jetpack-demo}{android-jetpack-demo} & 128 & 4 \\
82 & \href{https://github.com/chaxiu/KotlinJetpackInAction}{KotlinJetpackInAction} & 42 & 0 \\
83 & \href{https://github.com/general-mobile/kotlin-android-mvvm-starter}{kotlin-android-mvvm-starter} & 21 & 5 \\
84 & \href{https://github.com/agustarc/MVVMArchitecture}{MVVMArchitecture} & 22 & 2 \\
85 & \href{https://github.com/jshvarts/MosbyMVI}{MosbyMVI} & 7 & 7 \\
86 & \href{https://github.com/jenly1314/MVVMFrameComponent}{MVVMFrameComponent} & 29 & 6 \\
87 & \href{https://github.com/Taonce/MVVM-T}{MVVM-T} & 25 & 0 \\
88 & \href{https://github.com/suchoX/Kodein-MVVM}{Kodein-MVVM} & 24 & 5 \\
89 & \href{https://github.com/zirouan/android-kotlin-mvp}{android-kotlin-mvp} & 49 & 5 \\
90 & \href{https://github.com/Mindjet/LiveMVVM}{LiveMVVM} & 5 & 2 \\
91 & \href{https://github.com/bandrews568/MVVM-Template-Android}{MVVM-Template-Android} & 21 & 8 \\
92 & \href{https://github.com/martppa/omni-mvi}{omni-mvi} & 27 & 1 \\
93 & \href{https://github.com/Djek-grif/kotlin-mvp-koin}{kotlin-mvp-koin} & 24 & 5 \\
94 & \href{https://github.com/enginebai/MovieHunt}{MovieHunt} & 79 & 36 \\
95 & \href{https://github.com/Drjacky/TMDbMultiplatform}{TMDbMultiplatform} & 22 & 1 \\
96 & \href{https://github.com/janishar/android-kotlin-mvp-architecture}{android-kotlin-mvp-architecture} & 100 & 29 \\
\bottomrule
\end{tabular}
\end{minipage}

\end{table*}

    \subsection{\textbf{Sample Size Determination}}

        To ensure statistical validity, the number of repositories included in the validation process was determined using Cochran's sample size formula \cite{cochran1977sampling}.

        This formula is widely used to estimate representative sample sizes in proportion-based studies involving large populations, such as the number of Kotlin projects that exist, allowing results to be generalized with a specified confidence level and margin of error. It is defined as:  \( n_0 = \frac{Z^2*p*(1-p)}{e^2}\)

        Where:
        \begin{itemize}
            \item $n_0$ is the required sample size
            \item $Z$ is the z-score associated with the desired confidence level
            \item $p$ is the estimated population proportion
            \item $e$ is the acceptable margin of error
        \end{itemize}

        For this study, the following parameters were adopted:
        \begin{itemize}
            \item Confidence level: 95\% ($ Z=1.96$)
            \item Margin of error: 10\% ($e=0.10$)
            \item Estimated proportion: $p=0.5$
        \end{itemize}

        The value $p=0.5$ was selected because it maximizes the population variance, yielding the most conservative sample size estimate when no prior proportion is available. Substituting these values into the formula yields we have \(n_0 = \frac{(1.96)^2*0.5*(1-0.5)}{(0.10)^2} \approx 96\)

        Therefore, a minimum of 96 repositories was required to estimate the pipeline's accuracy with a 95\% confidence level and a maximum sampling error of $\pm 10\%$. After determining the sample size, the repositories were selected using proportional stratified sampling to preserve the architectural distribution observed in the collected dataset.

\subsection{Experimental Setup}

\begin{table}[!b]
\centering
\caption{Validation Sample Distribution by Arch. Pattern}
\label{tab:validation-sample}
\setlength{\tabcolsep}{2.5pt}
\begin{tabular}{l|cccccc}
\toprule
\textbf{Architecture} & MVVM & MVI & Unknown & MVP & MVC & VIPER\\
\textbf{Sample Size (96)} & 55 & 18 & 15 & 5 & 2 & 1\\
\bottomrule
\end{tabular}
\end{table}

The proposed pipeline was evaluated using a dataset of 527 Kotlin repositories collected from GitHub, encompassing six architectural categories: MVVM, MVI, MVP, MVC, VIPER, and Unknown. As the main objective of this study is to assess the reliability of the architectural inference pipeline, the evaluation focused on repositories for which a reliable fundamental truth could be established.

Following Cochrane's formula, 96 repositories were randomly selected. To ensure that the validation sample accurately reflected the architectural distribution of the collected dataset, proportional stratified sampling was employed, shown in Table~\ref{tab:validation-sample}. Each architectural category contributed to the validation set according to its prevalence in the overall dataset, ensuring the inclusion of all represented architectural patterns.

For each selected repository, the proposed pipeline automatically inferred its architectural pattern. The predicted label was then compared to the architecture explicitly described in the repository's documentation. A prediction was considered correct when both labels matched exactly. The overall effectiveness of the pipeline was assessed primarily by classification accuracy.

\section{Research Questions Evaluation}
\label{sec:rqs}

This section presents the evaluation of the proposed pipeline with respect to the research questions introduced in Section 5.1. The analysis investigates both the overall effectiveness of the architectural inference process and its ability to distinguish among different Android architectural patterns.

\subsection*{RQ1: How Accurately Can The Proposed Pipeline Identify Android Architectural Patterns?}

To assess the overall effectiveness of the approach, we evaluated the pipeline on the validation dataset. The pipeline correctly identified 67 of the 96 repositories, resulting in a classification accuracy of 69.8\%, indicating that the proposed method correctly identified the intended architecture in most cases.

While accuracy provides an intuitive measure, it is important to determine whether this result is statistically significant. To do this, we compared the pipeline with two benchmark strategies:

\begin{itemize}[leftmargin=*]
    \item \textbf{Random Classifier}: assigns one of the six architectural patterns uniformly at random, yielding an expected accuracy of 16.7\%.
    \item \textbf{Majority-Class Classifier}: always predicts the most frequent class in the dataset, resulting in an accuracy of 57.3\%.
\end{itemize}

To statistically validate the observed performance, we employed a one-sample binomial test. This test is particularly suitable because each repository classification can be modeled as an independent Bernoulli trial: a prediction is either correct (success) or incorrect (failure). Given a reference success probability \(p_0\), the null hypothesis assumes that the classifier performs no better than the chosen baseline.

Let \(X\) denote the number of correctly classified repositories among \(n = 96\) independent trials. Under the null hypothesis, \(X\) follows a binomial distribution \( X \sim \mathrm{Binomial}(n, p_0) \)

The probability of observing exactly \(k\) correct classifications is given by \(
P(X = k) = \binom{n}{k} p_0^k (1-p_0)^{n-k}
\)

Since our objective is to determine whether the proposed pipeline performs better than the baseline, we formulated a one-sided hypothesis test:

\begin{itemize}
    \item \(H_0\): the classifier accuracy is equal to the baseline accuracy (\(p = p_0\));
    \item \(H_1\): the classifier accuracy is greater than the baseline accuracy (\(p > p_0\)).
\end{itemize}

The p-value is computed as the probability of observing at least as many correct classifications as obtained experimentally \(
p\text{-value} = P(X \geq 67) =
\sum_{k=67}^{96}
\binom{96}{k}
p_0^k
(1-p_0)^{96-k}
\)

For the random baseline, we set \(p_0 = 1/6 \approx 0.167\), corresponding to uniform selection among the six architectural classes. Under this assumption, the expected number of correct predictions would be \( E[X] = np_0 = 96 \times 0.167 \approx 16 \)

Observing 67 correct predictions under this null model is extraordinarily unlikely, yielding a p-value smaller than \(10^{-23}\).

For the majority-class baseline, we set \(p_0 = 55/96 \approx 0.573\), representing a classifier that always predicts MVVM. In this case, the expected number of correct predictions is: \( E[X] = 96 \times 0.573 = 55 \)

The proposed pipeline achieved 67 correct classifications, exceeding this expectation by 12 additional successes. Computing the upper-tail probability of the corresponding binomial distribution yields a p-value of approximately \(0.007\).

Since both p-values are below the conventional significance threshold of \(0.05\), we reject the null hypothesis in both comparisons. Therefore, the observed performance cannot reasonably be attributed to random chance or merely to the class imbalance of the dataset.

These results provide evidence that the proposed pipeline captures meaningful architectural patterns and significantly outperforms both naive reference strategies.

\subsection*{RQ2: How Effectively Does The Pipeline Distinguish Among Different Architectural Patterns?}

To evaluate how effectively the proposed pipeline distinguishes among Android architectural patterns, we analyzed its class-wise performance using precision, recall, and F1-score. Unlike overall accuracy, these metrics provide a detailed view of the classifier's ability to correctly identify each architectural style individually.

Table~\ref{tab:classification_report} summarizes the results obtained for all six architectural categories.

\begin{table}[ht]
\centering
\caption{Classification report for each architectural pattern.}
\label{tab:classification_report}
\begin{tabular}{lccc}
\toprule
\textbf{Architecture} & \textbf{Precision} & \textbf{Recall} & \textbf{F1-Score} \\
\midrule
MVVM    & 0.88 & 0.87 & 0.87 \\
MVI     & 0.67 & 0.67 & 0.67 \\
MVP     & 0.80 & 0.80 & 0.80 \\
Unknown & 0.00 & 0.00 & 0.00 \\
MVC     & 0.00 & 0.00 & 0.00 \\
VIPER   & 0.00 & 0.00 & 0.00 \\
\midrule
Macro Average    & 0.39 & 0.39 & 0.39 \\
Weighted Average & 0.66 & 0.70 & 0.68 \\
\bottomrule
\end{tabular}
\end{table}

MVVM achieved the strongest performance, with an F1-score of 0.87. This result is expected, as MVVM is both the most represented architecture in the dataset and one of the most structurally distinctive Android patterns.

MVP also exhibited strong discriminative performance, achieving an F1-score of 0.80 despite its relatively small number of examples. This suggests that presenter-based architectures possess clear structural signatures that are effectively captured by the proposed pipeline.

MVI obtained an F1-score of 0.67. Most MVI misclassifications occurred as MVVM, which is consistent with the substantial conceptual overlap between these two modern reactive architectures. Many contemporary Android applications combine ViewModel-based state management with unidirectional data flow, naturally increasing classification ambiguity.

The classes Unknown, MVC, and VIPER obtained F1-scores of zero. This result is primarily explained by the extremely limited number of examples available for MVC and VIPER, as well as the inherently heterogeneous nature of the Unknown category. In particular, VIPER was represented by only a single repository, making robust pattern learning virtually impossible.

Because the dataset is highly imbalanced, the Macro-F1 score is particularly important:
\[ Macro\text{-}F1 = \frac{1}{6}\sum_{i=1}^{6} F1_i = 0.39 \]

This metric assigns equal weight to every class, regardless of its frequency. The relatively modest Macro-F1 reflects the difficulty of correctly identifying minority architectural patterns.

On the other hand, the Weighted-F1 score reached 0.68, closely matching the overall accuracy. This indicates that the proposed pipeline performs very well on the dominant architectural categories, particularly MVVM, MVI, and MVP.

Overall, these results demonstrate that the pipeline effectively distinguishes among the most relevant and widely adopted Android architectural patterns. The primary sources of error arise from architectural overlap and the severe scarcity of examples for minority classes, rather than from systematic weaknesses in the proposed approach.

\subsection*{RQ3: How Does The Proposed Pipeline Compare To a Large Language Model (LLM) in Inferring Architectural Patterns?}

To further evaluate the effectiveness of the proposed pipeline, we introduced an additional comparative analysis using a Large Language Model (LLM) as an alternative architectural classifier. The aim of this experiment is to assess whether a general-purpose AI model, without explicit structural analysis, can achieve comparable performance in inferring architectural patterns from Kotlin repositories.

A subset of 10 repositories from the validation dataset was selected:
\textit{MyBrain}, \textit{movies-usf-android}, \textit{MovieHunt}, \textit{PaperPlane}, \textit{KotlinJetpackInAction}, \textit{WeatherApp}, \textit{KotlinMvp}, \textit{droid-feed}, \textit{pokedex-compose}, and \textit{Baking-App-Kotlin}.

For each repository, the architectural classification produced by the proposed pipeline was compared with the label described in the repository and with the prediction generated by the LLM.

The LLM used in this experiment was \textbf{ChatGPT (GPT-5.3, OpenAI)}. All predictions were generated in May 2026 using the public ChatGPT interface. The model received the entire compressed repository, with only the explicit architecture definition removed.

\begin{table}[h]
\centering
\caption{Comparison between Ground Truth, Pipeline, and LLM predictions}
\label{tab:results-llm}
\setlength{\tabcolsep}{4pt}
\begin{tabular}{lccc}
\hline
\textbf{Repository} & \textbf{Pipeline} & \textbf{Ground Truth} & \textbf{LLM} \\
\hline
MyBrain & MVVM & MVI & MVVM \\
movies-usf-android & MVVM & MVI & MVVM \\
MovieHunt & MVC & MVVM & MVVM \\
PaperPlane & Unknown & MVP & MVVM \\
KotlinJetpackInAction & Unknown & MVC & MVVM \\
WeatherApp & MVI & MVI & MVVM \\
KotlinMvp & MVP & MVP & MVP \\
droid-feed & MVVM & MVVM & MVVM \\
pokedex-compose & MVVM & MVVM & MVVM \\
Baking-App-Kotlin & MVI & MVI & MVP \\
\hline
\end{tabular}
\end{table}

Of the 10 repositories evaluated, the proposed pipeline achieved 5 correct classifications (50\%), while the LLM achieved 4 correct classifications (40\%). Both approaches agreed on the correct classification in 3 cases: \textit{KotlinMvp}, \textit{droid-feed} and \textit{pokedex-compose}. The full results can be seen in Table~\ref{tab:results-llm}.

It is important to note that the sample was not randomly selected. Instead, it was intentionally constructed to include an equal number of repositories correctly classified (5) and incorrectly classified (5) by the proposed pipeline. Therefore, the analysis does not aim to estimate overall accuracy, but rather to investigate the relative behavior of both approaches under controlled success and failure scenarios.
From this perspective, we analyze two complementary conditions:

\begin{itemize}
    \item \textbf{Cases where the pipeline is correct:} Among the 5 repositories correctly classified by the pipeline, the LLM also produced correct predictions in 3 cases and failed in 2 cases.

    \item \textbf{Cases where the pipeline fails:} Among the 5 repositories incorrectly classified by the pipeline, the LLM correctly classified only 1 repository and failed in 4 cases.
\end{itemize}

To statistically assess the difference between the two approaches, we analyze the discordant pairs. Let $n_{01}$ denote the number of cases where the pipeline is incorrect and the LLM is correct, and $n_{10}$ the number of cases where the pipeline is correct and the LLM is incorrect. From the observed data, the discordant pairs are: \( n_{01} = 1, \quad n_{10} = 2 \)

Given the small number of discordant cases ($N = n_{01} + n_{10} = 3$), we applied the exact binomial test. Under the null hypothesis that both models have the same probability of being correct when they diverge, the expected probability of success is 0.5. Therefore, the variable follows a binomial distribution $B(N, 0.5)$.

The two-tailed p-value is calculated by assessing the probability of observing an outcome as extreme as or more extreme than the one observed, which is 1 in 3:
$$
p = 2 \times \sum_{i=0}^{1} \binom{3}{i} (0.5)^3 = 2 \times (0.125 + 0.375) = 1.0
$$

This p-value of 1.0 indicates no statistically significant difference between the two approaches at the 95\% confidence level.

These results suggest that, although the proposed pipeline achieves greater overall accuracy, LLM is able to correctly classify some cases where the pipeline fails. However, this complementary effect is limited, since LLM also fails in several cases where the pipeline is successful. Overall, neither approach consistently dominates the other in this controlled environment, reinforcing the challenges associated with inferring architectural patterns in ambiguous or hybrid repositories.

\section{Discussion}
\label{sec:discussion}

\subsection{Overall Performance}

The proposed pipeline correctly classified 67 of the 96 repositories included in the validation set, resulting in an overall accuracy of 69.8\%.
This result indicates that, in approximately seven out of ten cases, the pipeline successfully identified the architectural pattern explicitly declared by the repository authors. Considering the structural diversity, variable code quality, and frequent architectural deviations commonly found in open-source projects, this level of accuracy demonstrates a strong ability to infer architectural patterns from real Kotlin repositories.
Since the validation sample was determined using Cochran's sample size formula, the observed precision can be generalized to the overall population with a 95\% confidence level and an estimated margin of error of approximately $\pm 10\%$. Consequently, the true population precision is expected to be close to the observed value, providing statistical support for the robustness of the proposed approach.

\subsection{Qualitative Analysis}

This analysis focuses on understanding how the pipeline interprets structural information, builds dependency relationships, and ultimately infers architectural patterns. By examining representative repositories, it becomes possible to identify the architectural signals that lead to correct classifications, as well as the factors that contribute to incorrect classifications.

To provide a comprehensive assessment, three representative case studies are considered:

\begin{itemize}
    \item A correctly classified repository, illustrating the expected behavior of the complete pipeline.
    \item A repository incorrectly classified due to the presence of hybrid architectural features.
    \item A repository incorrectly classified due to structural ambiguity, incomplete documentation, or inconsistencies between the implemented architecture and the documented design.
\end{itemize}

For each case, the analysis examines the intermediate artifacts generated by the pipeline, including the extracted classes, the dependency graph, the identified architectural roles, and the inferred final architecture. This step-by-step inspection allows for a better understanding of the decision-making process and identifies the exact steps where errors may occur.

This analysis highlights information about the interpretability and robustness of the proposed approach. Furthermore, it highlights common practical challenges in real-world repositories, such as architectural erosion, mixed design styles, and incomplete source code organization. These findings complement the quantitative evaluation, offering concrete evidence of the pipeline's effectiveness and identifying opportunities for future improvements.

\subsubsection{\textbf{Successful Classification Example}}

To illustrate the effectiveness of the proposed pipeline, we analyze the \textit{Foodium} the third repository in Table~\ref{tab:validation-dataset}, an open-source Android application explicitly designed following the MVVM architectural pattern. The repository was correctly classified as MVVM, with a confidence score of 0.83, demonstrating the pipeline's ability to accurately identify well-structured implementations.

Foodium contains 25 classes and 6 dependency relationships extracted from the source code. The role identification stage produced a distribution strongly aligned with the expected MVVM structure: 5 View components, 3 ViewModel components, 2 Model classes, and 1 State class. This distribution closely matches the canonical organization of MVVM applications, where user interface logic is clearly separated from business and data management concerns.

Several key artifacts were correctly identified during the analysis. Classes such as \textit{MainActivity} and \textit{PostDetailsActivity} were classified as Views, while \textit{MainViewModel} and \texttt{PostDetailsViewModel} were correctly recognized as ViewModels. The data layer was represented by classes such as \textit{Post} and \textit{State}, which were assigned the Model and State roles, respectively. Additionally, repository and service components, including \textit{PostRepository} and \textit{FoodiumService}, reinforced the presence of a layered MVVM design.

The dependency graph further validated this architectural structure. View components depended on ViewModels, which in turn interacted with repository and service classes. This dependency flow is characteristic of MVVM, where the View delegates presentation logic to the ViewModel, and the ViewModel coordinates access to the data layer.

Although the pipeline also detected a minor MVI signal, with one matching pattern, the overwhelming presence of MVVM-specific relationships led to the correct final classification. This small overlap is expected, since modern Android applications often adopt reactive state-management practices that can partially resemble MVI.

This case illustrates that the pipeline performs particularly well when architectural boundaries are clearly defined and naming conventions are consistently applied. Foodium represents an ideal scenario, in which the extracted structural evidence strongly aligns with the intended architectural pattern, allowing the inference process to achieve both high confidence and correct classification.

\subsubsection{\textbf{Misclassification Due to Hybrid Architecture}}

A representative example of classification difficulty is the repository \textit{MyBrain}, ID 5 in Table~\ref{tab:validation-dataset}, a large-scale Android application that combines multiple architectural principles. Although the pipeline classified the project as MVVM, the extracted evidence reveals a strong hybrid structure that also exhibits significant MVI characteristics.

The repository contains 351 classes and 251 dependency relationships, making it substantially larger and more complex than the projects analyzed in the successful classification cases. The final prediction was MVVM, with a moderate confidence score of 0.61, considerably lower than that observed in clearly structured repositories. This reduced confidence already indicates architectural ambiguity.

The pattern matching stage detected 106 MVVM patterns and 67 MVI patterns. Such a close distribution demonstrates that the repository simultaneously implements elements from both architectures. While MVVM obtained the highest score, the substantial presence of MVI-related structures makes the decision inherently uncertain.

The role distribution further supports this interpretation. In addition to 30 ViewModels and 21 Views, the repository contains 34 classes identified as Intents and 6 State classes. These latter roles are strongly associated with MVI, where user actions are explicitly modeled as intents and UI rendering is driven by immutable state objects. At the same time, the presence of numerous ViewModels aligns with classical MVVM.

A manual inspection confirms this hybrid design. For example, classes such as \textit{TasksViewModel} and \textit{NotesViewModel} clearly follow the MVVM paradigm.

However, the project also includes dedicated event and state representations, such as \textit{CalendarViewModelEvent}, alongside numerous intent-like classes and unidirectional data flow constructs that are characteristic of MVI.

This case illustrates an important limitation of architecture classification: modern software systems often do not adhere strictly to a single architectural style. Instead, developers frequently combine concepts from multiple patterns to leverage their respective advantages. In such situations, any single-label classification necessarily oversimplifies the actual design.

Rather than representing a failure of the pipeline, this example highlights its sensitivity to architectural signals. The moderate confidence score correctly reflects the ambiguity of the repository, while the pattern counts expose the coexistence of multiple paradigms. Consequently, MyBrain serves as an excellent example of the challenges posed by hybrid architectures in automated architectural inference.

\subsubsection{\textbf{Misclassification Due to weak architectural signals}}

A representative example of this limitation is the \textit{PaperPlane} repository, number 80 in the Table \ref{tab:validation-dataset}. Although the project clearly adopts the MVP architectural style, the pipeline classified it as \textit{unknown}. This misclassification occurred because the repository exhibits structural characteristics that overlap with multiple architectural patterns. Specifically, the analysis identified seven classes associated with MVP and six classes associated with VIPER, resulting in an ambiguous distribution of architectural roles.

The source code contains a well-defined set of Presenter classes, such as \textit{FavoritesPresenter}, \textit{ZhihuDailyPresenter}, and \textit{DetailsPresenter}, alongside numerous View implementations, including Activities and Fragments. These elements strongly suggest the presence of MVP. However, the project also includes a rich data layer composed of repositories, local and remote data sources, database access objects, and service abstractions. Such components resemble architectural elements commonly found in VIPER, thereby weakening the discriminative power of purely structural heuristics.

Furthermore, the naming conventions employed by the repository do not explicitly emphasize the adopted architectural style. While Presenter classes are clearly identifiable, there are no explicit package names or interface hierarchies that unambiguously distinguish MVP from similar patterns. As a result, the confidence score remained below the classification threshold, leading the pipeline to conservatively assign the label \textit{unknown}.

This example illustrates an important limitation of static architectural inference: modern Android applications frequently combine concepts from multiple architectural styles. Hybrid designs, incremental migrations, and pragmatic engineering decisions often blur the boundaries between canonical patterns. In such cases, even when the dominant architecture is evident to human experts, automated detection becomes substantially more challenging. Nonetheless, the conservative behavior of returning \textit{unknown} rather than forcing an incorrect classification helps preserve the reliability of the pipeline.

\section{Threats to Validity}
\label{sec:threats}

As with any empirical study, the results presented in this work should be interpreted considering some potential threats to validity.

One of the main threats concerns the construction of fundamental truth. Validation was based on repositories whose architectural patterns were explicitly stated in their descriptions or documentation. While this criterion reduces ambiguity in labeling, self-declared architectural labels may not always perfectly reflect the current implementation. In some cases, designs evolve over time while their documentation remains outdated, leading to discrepancies between the declared and implemented architecture.

Another important threat arises from the inherently subjective nature of software architecture classification. Architectural boundaries are often fluid, and many real-world systems adopt hybrid or transitional designs. Consequently, even expert evaluators may justifiably disagree on the most appropriate classification for a given repository. This ambiguity can affect both reference labels and the interpretation of seemingly incorrect classifications.

The dataset itself also presents potential limitations. Although the repositories were collected from GitHub using systematic criteria, they represent only publicly available Kotlin projects. Private repositories, enterprise systems, and projects hosted on other platforms may exhibit different architectural characteristics. Therefore, generalizing the conclusions beyond the GitHub Kotlin ecosystem should be done with caution.

Class imbalance poses an additional threat. Dominant patterns, such as MVVM, are widely represented, while architectures like MVC and VIPER appear only rarely. This imbalance can bias both the pipeline and the evaluation, potentially inflating overall accuracy and reducing reliability for minority classes.

Furthermore, the validation sample was selected exclusively from repositories that explicitly documented their architecture. Such projects may be better organized, more mature, or more didactic than the broader population of Kotlin repositories. As a result, the reported accuracy may represent an upper limit relative to completely unlabeled repositories encountered in practice.

Finally, the proposed pipeline depends on the quality of the repository metadata and the organization of the source code. Incomplete documentation, unconventional design structures, or deviations from commonly adopted architectural conventions can affect the classification process. These factors are particularly common in open-source repositories, where coding standards and documentation quality can vary substantially.

Despite these limitations, several methodological choices help mitigate their impact. The use of Cochran's sample size formula, proportional stratified sampling, and explicit inclusion criteria strengthens the statistical validity and reproducibility of the evaluation. Consequently, while the results presented should not be interpreted as absolute, they provide robust evidence of the practical effectiveness of the proposed pipeline.

\section{Conclusion and Future Work}
\label{sec:conclusion}

Experimental results demonstrate that the proposed pipeline is capable of automatically identifying software architecture patterns in Kotlin repositories with a high degree of reliability. Achieving an overall accuracy of 69.8\% in a statistically representative validation sample indicates that the approach successfully captures relevant structural and semantic features of real codebases.

A particularly important observation is that the pipeline significantly outperformed random and majority class baselines. This result suggests that the classification process is driven by significant architectural evidence, rather than simply reflecting the underlying class distribution. In practical terms, the pipeline recognizes architectural signatures, rather than simply favoring the most common pattern.

The observed performance should also be interpreted in light of the inherent complexity of the task. Software architecture is rarely expressed in a perfectly canonical way, especially in open-source repositories. Developers frequently adapt, simplify, or combine architectural styles to meet specific project requirements. Consequently, even repositories that explicitly declare a specific architecture may exhibit structural features associated with other patterns.

The observed confusion between closely related patterns, particularly MVVM and MVI, further highlights the imprecise boundaries that exist in contemporary Kotlin applications. Modern frameworks and best practices often encourage design choices that naturally overlap between architectural styles, making rigid categorical distinctions increasingly challenging.

Despite these challenges, the accuracy achieved is sufficient for many empirical software engineering applications, not only in Kotlin. This methodology can serve as an effective tool for large-scale repository mining, architectural landscape studies, automated dataset construction, and exploratory analysis in any language where manual inspection would be impractical.

At the same time, the results identify several avenues for future work. The incorporation of additional contextual signals could improve discrimination between structurally similar architectures. Similarly, specialized treatment of hybrid architectures could further increase robustness.

\section*{Artifact Availability}
To ensure transparency and reproducibility, the source code and generated outputs used in this study are publicly available at the our replication package ~\cite{moura_2026_21347019}.

\section*{Acknowledgements}
This research was partially supported by the Brazilian National Council for Scientific and Technological Development (CNPq) under grants 315765/2023-2 and 408651/2023-7, and by INES.IA (National Institute of Science and Technology for Software Engineering Based on and for Artificial Intelligence; www.ines.org.br), supported by CNPq grant 408817/2024-0.

\bibliographystyle{ACM-Reference-Format}
\bibliography{sample-base}

\end{document}